\documentclass[12pt]{article}

\usepackage{amsfonts}

\begin{document}

\title{Biology is a constructive physics}

\author{S.V. Kozyrev}

\maketitle

\centerline{\it Steklov Mathematical Institute of Russian Academy of Sciences, Moscow, Russia}

\begin{abstract}
Yuri Manin's approach to Zipf's law (Kolmogorov complexity as energy) is applied to investigation of biological evolution.
Model of constructive statistical mechanics where complexity is a contribution to energy is proposed to model genomics.
Scaling laws in genomics are discussed in relation to Zipf's law. This gives a model of Eugene Koonin's Third Evolutionary Synthesis --- physical model which should describe scaling in genomics.
\end{abstract}

\section{Introduction}

In the present paper we discuss application of Yuri Manin's idea \cite{Manin}, \cite{Manin2} on relation of the Zipf's law (power law distribution for frequencies of words in texts) and Kolmogorov order (complexity as energy) to biological evolution.

The standard Darwinian approach to evolution is to consider evolution as a result of heredity, mutations and selection. A simplest mathematical model for Darwinian evolution is a model of random walk on the landscape of biological fitness.
We propose to consider a model of random walk on a landscape, where the space is constructive  --- evolution runs in the space of sequences (genomes) generated by a set of constructive operations (genome editing operations). This set of computable genome editing operations generates the corresponding Kolmogorov complexity and Kolmogorov order in the space of sequences.  Moreover, in the model under consideration the Hamiltonian of evolution (used for the definition of the random walk) should contain a contribution given by (estimate from above for) Kolmogorov complexity. This contribution (weighted sum of scores of evolutionary transformations) describes evolutionary effort to generate a genome and expresses the idea that simpler genomes are more advantageous.

Zipf's law in this approach should be related to power laws observed in genomics. It was proposed by Eugene Koonin \cite{Koonin1}, \cite{Koonin2} to develop a statistical mechanical model of evolution, where genetic ensemble should be described by interacting gas of genes (the Third Evolutionary Synthesis). This model should describe several scaling laws for gene frequencies in genomes observed in genomic data. In our approach these scaling laws should be a result of complexity contribution in the Hamiltonian of biological evolution and should be explained by the same mechanism as the Zipf's law.

In general this approach expresses the idea that biological objects are designed (by the process of evolution) and biological structures and functions can be considered from algorithmic point of view. Therefore some biological observations, in particular scaling laws in genomics, should express properties of world of algorithms (or Platonic world of ideas), in particular properties of Kolmogorov complexity.

Approach of the present paper (biological evolution in a constructive world) can be compared with Code Biology \cite{Barbieri} which investigates  multiplicity of codes in biological systems.
These codes include genetic code: amino acid code, genomes;
epigenetic codes: histone code, methylation code, splicing codes, chromatin codes;
codes in multicellular organisms and bacterial colonies;
codes in neural sciences. Discussion of the Code Biology and different biological codes can be found in different papers in the issue of BioSystems \cite{Barbieri}.

The following properties of codes were discussed in code biology:

A code is a mapping between objects of two worlds implemented by objects of a third world.

A code is a pattern in a sequence which corresponds to some specific biological function.

The set of maps (codes) should be rich enough to generate a rich set of biological functions.

Codes are generated in the process of evolution --- more complex organisms use more codes.

We will use the notion of Constructive World as a mathematical model of Code Biology.

Scaling laws in biological models were discussed in many papers, in particular, scaling in biology was discussed applying random networks \cite{ScaleFree}, \cite{Nechaev}.
Scaling in protein dynamics and relation to theory of disordered systems (spin glasses) was discussed in particular in \cite{Ansari}, \cite{Frauenfelder}, \cite{ABK}.
For recent discussion of evolution in analogy to spin glasses see \cite{Katsnelson}. For investigation of disordered systems (in particular proteins) ultrametric and $p$-adic methods were applied \cite{ABK}, \cite{PaSu}, \cite{tmf2014_1}. For application of $p$-adic methods to modeling of genetic code see \cite{Plane}, \cite{Branko}.

Another useful point of view in theory of biological evolution is related to machine learning. A problem of biological evolution can be considered as a problem of learning where genomes learn in the process of natural selection. Biological fitness in this approach can be compared to the functional of empirical risk (number of errors of classifier related to a genome on a training set) and Kolmogorov complexity of a genome (evolutionary effort to generate a genome) plays a role of regularizer to reduce overfitting in a problem of learning by evolution.

\section{Complexity as energy}

In the present section we discuss, following Yu.I.Manin \cite{Manin}, the relation of Kolmogorov complexity and Zipf's law. The general statement of \cite{Manin} is:

\medskip

{\sl There are natural observable and measureable phenomena in the world of information that can be given a mathematical explanation, if one postulates that logarithmic Kolmogorov complexity plays a role of energy.}

\medskip

We apply this idea to modeling of biological phenomena and discuss relation of Zipf's law and scaling in genomes. Our approach is based on observation that
biological (genetic) sequences form a constructive world, i.e. {\it biological objects are designed}. In this sense biology is a constructive physics.

In general the idea of constructive mathematics is to add computability to properties of mathematical theory, in particular to consider theories where objects are defined constructively and maps are computable functions.

\medskip

\noindent{\bf Constructive worlds.}\quad
An (infinite) constructive world is a countable set $X$ given together with a class of structural numberings: computable bijections $u: \mathbb{Z}_+ \to X$.
Moreover natural maps between constructive worlds should be given by computable functions.

\medskip

\noindent{\bf Example}. Goedel numbering of formulae of a formal language.

\medskip

\noindent{\bf Kolmogorov complexity and Kolmogorov order} \cite{Kolmogorov1965}, \cite{Kolmogorov1969}, \cite{Manin1}.\quad
Let $X$ be a constructive world. For any (semi)-computable function $u:\mathbb{Z}_+\to X$, the
(exponential) complexity of an object $x\in X$ relative to $u$ is
$$
K_u(x) = \min \{m\in\mathbb{Z}_+| u(m)=x \}.
$$

If such $m$ does not exist, we put $K_u(x) = \infty$.

\medskip

\noindent{\bf Claim}.\quad{\it  There exists such $u$ (an optimal Kolmogorov numbering) that
for each other (semi)-computable $v:\mathbb{Z}_+\to X$, some constant $c_{uv} > 0$, and all $x \in X$, one has
$$
K_u(x) \le c_{uv} K_v(x).
$$
}

This $K_u(x)$ is called (exponential) Kolmogorov complexity of $x$.

\medskip

A {\bf Kolmogorov order} in a constructive world $X$ is a bijection $X\to \mathbb{Z}_+$ arranging
elements of $X$ in the increasing order of their complexities $K_u$.

\medskip

\noindent{\bf Logarithmic Kolmogorov complexity}.\quad
Let us consider a set $X$ (a constructive world) generated by ''programs'' (sequences of bits).
For $x,y\in X$ we consider the conditional entropy (complexity) as the minimal length of program $p$ (in bits) satisfying
\begin{equation}\label{K(|)}
K_A(x|y)=\min_{A(p,y)=x} l(p).
\end{equation}
i.e. program $p$ computes $x$ starting from  $y$. Here $A$ is a ''way of programming''.

Unconditional complexity is given by application of the above definition to some ''initial'' object $y_0$
\begin{equation}\label{K()}
K_A(x)=K_A(x|y_0).
\end{equation}

Logarithmic Kolmogorov complexity of $x$ is the length (in bits) of the shortest program which generates $x$ (logarithmic complexity is close to logarithm of exponential complexity).
In particular, there exists such way of programming $A$ that
for each other (semi)-computable $B$, some constant $c_{AB} > 0$, and all $x \in X$, one has
$$
K_A(x) \le K_B(x) + c_{AB}.
$$

Here $K_A(x)$ is the logarithmic Kolmogorov complexity of $x$.

\medskip

\noindent{\bf Properties of Kolmogorov complexity}.

Any optimal numbering is only partial function, and its definition domain is not decidable.

Kolmogorov complexity $K_u$ itself is not computable. It is the lower bound of a
sequence of computable functions. Kolmogorov order is not computable as well.

Kolmogorov order of naturals cardinally differs from the natural order in the following
sense: it puts in the initial segments very large numbers that are at the same
time Kolmogorov simple (for example $2^k$, $2^{2^k}$).

This can be compared with properties of natural language, which are usually discussed as a result of historical accidents but at least partially are related to possibility to express complex meanings in short way  --- abundance of synonyms and senseless grammatically correct texts.

\medskip

\noindent{\bf Zipf's law and Kolmogorov order}.\quad Zipf's law describes frequencies of words of a natural language in texts. If all words $w_k$ of a language are ranked according to decreasing frequency of their appearance in a corpus of texts, then the frequency $p_k$ of $w_k$ is approximately inversely proportional to its rank $k$: $p_k\sim k^{-1}$.
Zipf: this distribution ''minimizes effort''.

\medskip

\noindent{\bf How minimization of complexity leads to Zipf's law} \cite{Manin}.\quad
It is easy to see that Gibbs distribution with energy proportional to $\log k$ gives a power law.

A mathematical model of Zipf's law is based upon two postulates:

(A) Rank ordering coincides with the Kolmogorov ordering.

(B) The probability distribution producing Zipf's law is $K(w)^{-1}$.

Thus Zipf's law is described by the statistical sum
$$
\sum_{w}K(w)^{-z}
$$
with the inverse temperature $z=1$.

For natural numbers, since for majority of naturals exponential Kolmogorov complexity of $n$ is close to $n$, the statistical sum is ''similar'' to zeta function
\begin{equation}\label{zeta}
\sum_{n}K(n)^{-z}\approx\zeta(z)=\sum_{n} n^{-z}.
\end{equation}

Point $z=1$ is the point of phase transition: for $z>1$ the series for zeta function converge and for $z\le 1$ diverge.
Exponent $-1$ in the Zipf's law can be discussed in relation to  phase transition in the model where (logarithmic) Kolmogorov complexity is energy.

\section{Zipf's law and scaling in genomics}

In the present section we introduce a model of biological evolution based on constructive statistical mechanics and discuss relation of scaling in genomics and Manin's approach to Zipf's law exposed in the previous section.

Discussion of universal properties of genome evolution can be found in \cite{Koonin1}, \cite{Koonin2}. The following universal statistical properties were observed in genomics:

1) log--normal distribution of the evolutionary rates between orthologous genes;

2) power law--like distributions of membership in paralogous gene families;

3) scaling of functional classes of genes with genome size.

At least some of these observations can be compared to Zipf's law.

Eugene Koonin proposed to develop a statistical mechanical model of biological evolution based on gene ensembles. This hypothetical model was based on the idea about interacting gas of genes and called ''the third evolutionary synthesis'' (the first is Darwinism, the second is Darwinism plus genetics, and the third should generalize Darwinism with genomic data). In particular  \cite{Koonin2}:

\medskip

{\sl The universality of several simple patterns of genome and molecular phenome evolution, and the ability of simple mathematical models to explain these universals,
suggest that ''laws of evolutionary biology'' comparable in status to laws of physics might be attainable.}

\medskip

Here we discuss scaling in genomics as a result of presence of contribution from complexity in energy. Complexity of a sequence describes the evolutionary effort to generate a sequence.

\medskip

\noindent{\bf Weighted logarithmic complexity as evolutionary effort}.\quad Let us consider a structure of constructive world in the set of genomic sequences.
We start with a finite set $S$ of sequences (genes, regulatory sequences, etc.), and a finite set $O$ of genome editing operations with contains operations of gluing together sequences and operations similar to typical evolutionary transformations (point mutations, insertions, deletions (in particular insertions and deletions of genes $s_i\in S$), duplications of parts of a sequence, etc.).

To elements $s_i\in S$ and $o_j\in O$ we put in correspondence positive numbers $w(s_i)$ and $w(o_j)$, called scores (or weights), and to any sequence $s$ obtained from elements in $S$ by applications of operations in $O$ we put in correspondence score $w(s)$ equal to a sum of scores of elements $s_i\in S$ and operations $o_j\in O$: composition of $o_j$ generates the sequence $s$ starting from $s_i$.

A sequence $s$ can be obtained in this way non-uniquely  --- in particular, one can insert a symbol at a given position and then delete this symbol leaving $s$ unchanged.
Let us define complexity of $s$ as a minimum over possible compositions of operations $o_j\in O$ and elementary sequences $s_i\in S$ giving $s$
\begin{equation}\label{K_SOW}
K_{SOW}(s)=\min_{A(s_1,\dots,s_n)=s}\left[\sum_{i}w(s_i)+\sum_{j}w(o_j)\right]
\end{equation}
where $A$ is a (finite) composition of $o_j$ applied to sequences $s_1,\dots,s_n$.

Complexity $K_{SOW}(s)$ can be considered as a weighted number of genes and edit operations generating sequence $s$. Logarithmic complexity can be (approximately) discussed as a number of edit operations generating element of a constructive world, the above complexity (\ref{K_SOW}) gives weighted version of estimate from above for logarithmic Kolmogorov complexity (in particular this estimate from above should be computable).

Conditional version $K_{SOW}(s'|s)$ of weighted complexity (\ref{K_SOW}) has the form
\begin{equation}\label{K_SOW1}
K_{SOW}(s'|s)=\min_{A(s_1,\dots,s_n)[s]=s'}\left[\sum_{i}w(s_i)+\sum_{j}w(o_j)\right],
\end{equation}
where we generate sequence $s'$ starting from sequence $s$. Here $A(s_1,\dots,s_n)[s]$ is a combination of genome editing operations containing sequences $s_1, \dots, s_n$  applied to $s$.

\medskip

In the case when the set of edit operations contains insertions, deletions and substitutions of symbols the weighted conditional complexity $K_{SOW}(s'|s)$ reduces to the alignment score (and the program which transforms $s$ to $s'$ is alignment, see \cite{Pevzner} for discussion of alignment of sequences).

\medskip

\noindent{\bf Complexity as energy in biological evolution}. \quad
Let us consider a statistical mechanical system where states $s$ are sequences, generated as above, and the statistical sum is equal to
\begin{equation}\label{Z}
Z=\sum_{s}e^{-\beta H(s)},\quad H=H_F+H_K
\end{equation}
where $\beta$ is the inverse temperature and the Hamiltonian contains two contributions, the contribution $H_F(s)$ describes biological fitness of the sequence $s$ and the contribution $H_K(s)$ describes complexity of $s$ (in particular, for the described above example $H_K=K_{SOW}$). Here good fitness corresponds to low $H_F(s)$ (potential wells on the fitness landscape). The symbol $K$ in $H_K$ is for Kolmogorov (complexity).

This statistical sum describes a model of constructive statistical mechanics (since we have a constructive procedure of generation of sequences).
We consider this constructive statistical system as a model of biological evolution.
The contribution $H_K(s)$ in the energy describes the evolutionary effort to generate sequence $s$ (sequences with less evolutionary effort are more advantageous).

\medskip

\noindent{\bf Remark 1}.\quad Since by our construction the complexity of sequences grow sufficiently fast with addition of sequences from $S$ and application of editing operations from $O$, for sufficiently low temperatures (large $\beta$) the constructive statistical sum (\ref{Z}) converges. Let us note that if we would enumerate sequences by lexicographical order and would not take into account the complexity term in the Hamiltonian, we would obtain for the statistical sum a divergent expression which does not make sense.

\medskip

\noindent{\bf Remark 2}.\quad Discussed in relation to Zipf's law ''abundance of synonyms and senseless grammatically correct texts''
can be also applied for genomic sequences --- senseless texts are similar to junk DNA, synonyms and possible different meanings of texts can be discussed as complex relations between genotype and phenotype.

\medskip

\noindent{\bf Remark 3}.\quad Using (\ref{K_SOW}), (\ref{Z}), the scaling in genomics, in particular power law--like distributions of membership in paralogous gene families, can be discussed as follows.  Let us recall that paralogous genes are genes in the same genome generated by duplication events.

Let us assume that if the genome contains a paralogous family of genes with $N$ elements, the Kolmogorov rank of this genome should be proportional to $N$ (since the complexity (\ref{K_SOW}) in this case will contain $N$ contributions $w(s_i)$ for some gene $s_i$). Then by Zipf's law (\ref{zeta}) contribution of this genome to the statistical sum will be proportional to $N^{-z}$ which gives the power law.

\section{Evolution and machine learning}

Application of ideas of machine learning in biology is a natural approach. A problem of biological evolution can be considered as a problem of learning where genomes learn in the process of natural selection.

\medskip

\noindent{\bf Machine learning}.\quad
Problem of learning of a classifier is a joint minimization of empirical risk and regularization term
$$
R_{\rm emp}({\rm classifier},{\rm training~ set})+Reg({\rm classifier}).
$$

Functional of empirical risk measures number of errors on a training set
$$
R_{\rm emp}(s)={1\over l}\sum_{j=1}^{l}(y_j-f(v_j,s))^2.
$$
Here $s$ parameterize classifiers $f(v,s)$. Classifier $f(v,s)$  classifies a situation $v$ (gives 0 or 1 for $f(v,s)$). Here $(y_1,v_1)$, \dots, $(y_l,v_l)$, $y_j\in{0,1}$ is the training set.

Regularizing contribution describes some kind of complexity of a classifier. Minimization of this term reduces overfitting (memorization of training sets which give classifiers with low generalization power).
In particular in Vapnik--Chervonenkis theory (or VC-theory) \cite{Vapnik} a classifier can be taught if the family of classifiers has sufficiently low VC-entropy (which is some kind of complexity).

\medskip

In Hamiltonian (\ref{Z}) of biological evolution $H=H_F+H_K$ the contribution $H_F(s)$  (biological fitness of sequence $s$) can be modeled by the functional of empirical risk
$H_F(s)=R_{\rm emp}(s)$ (genome $s$ recognizes situation $v$ using classifier of biological function $f(v,s)$, empirical risk measures the pressure of selection) and the complexity $H_K$ (evolutionary effort) is the regularization term. In this sense minimization of evolutionary effort reduces overfitting in biological evolution.

Minimization of Kolmogorov complexity can be considered as a regularization suitable for generative models in constructive worlds.
In particular this can be important for deep learning where deep learning networks are considerably simpler (contain less nodes) in comparison to shallow learning networks.

\section{Conditional complexity and hyperbolic geometry}

Let us consider a structure of oriented graph on the defined above constructive world of sequences. Sequences are vertices of the graph, sequences $s$ and $s'$ are connected by oriented edge $ss'$ if there exists transformation in $O$  which maps $s$ to $s'$ (or there exists sequence $s_i\in S$ which maps $s$ to $s'$ if we glue $s_i$ to $s$). Scores of edit operations define weights of edges of the oriented graph of sequences.

Let us define a distance function from $s$ to $s'$ to be given by conditional complexity (\ref{K_SOW1}). This distance function on the graph of sequences is not a metric (in particular, it is not necessarily symmetric). This function can be discussed as weighted edit distance, or weighted number of edit (evolutionary) transformations which convert $s$ to $s'$.
If the set of edit transformations is sufficiently rich, with this distance function the space of sequences will be hyperbolic --- volumes of balls will grow exponentially with the radii.

The distance function (\ref{K_SOW1}) is non-symmetric because evolution is irreversible --- if for any oriented edge $ss'$ (corresponding to evolutionary event) of the graph of sequences would exist the inverse edge $s's$ with the same score, distance function (\ref{K_SOW1}) would be symmetric (it will define the standard weighted distance on a graph).

Hyperbolic geometry can be discussed in relation to Zipf's law. Zipf's law was given by Gibbs distribution on natural numbers where energy was proportional to logarithm. Approximately logarithmic dependence of Kolmogorov complexity for majority of natural numbers translates to exponential dependence of volume of the set of naturals with limited logarithmic complexity on the upper bound of complexity.

\medskip

Let us discuss the optimization problem in hyperbolic spaces:

\smallskip

--- We consider energy landscape on a hyperbolic space with potential wells of limited depth.

\smallskip

--- We perform optimization in a potential well of the landscape.

\smallskip

--- To perform optimization in short time a potential well should have small radius (to perform optimization in limited number of steps).

\smallskip

--- Since balls are hyperbolic we can perform search over exponentially large volume hidden inside a potential well of limited radius (if a potential well has funnel--like shape with global minimum and sufficiently large gradient).

\medskip

Our conjecture is that biological evolution works because optimization in hyperbolic spaces is more effective.
Actually similar idea was proposed for explanation of protein folding (folding funnels) \cite{funnel}.

\medskip

\noindent{\bf Summary}.\quad We have discussed the application of Yuri Manin's idea on relation of Zipf's law and Kolmogorov order {\it (complexity as energy)} to biological evolution --- the Hamiltonian of evolution should contain a contribution given by (estimate from above for) Kolmogorov complexity, the computable enumeration of genetic sequences is given by a variant of Kolmogorov order for combinations of elementary evolutionary operations.

Zipf's law in this approach should be related to power laws observed in genomics, in this sense complexity as energy approach is related to the proposal by Eugene Koonin to develop {\it third evolutionary synthesis} in analogy to physical laws, as a model of statistical mechanics which describes interacting gas of genes.

Approach of the present paper expresses the idea that {\it biological objects are designed} (by the process of evolution) and biological structures and functions can be considered from algorithmic point of view.
We discuss the analogy between the model of evolution and machine learning where selection pressure takes the form of the functional of empirical risk and (estimate for) Kolmogorov complexity (evolutionary effort) gives the regularization term which reduces overfitting in evolution.

\medskip

\noindent{\bf Acknowledgments.}
This work is supported by the Russian Science Foundation under grant 14--50--00005.

\end{document}